\documentstyle[11pt,epsf]{article}

\newcommand{\sign}{\mbox{\rm{sign}}}

\newcommand{\brmP}{\mbox{\bf{P}}}
\newcommand{\brmI}{\mbox{\bf{I}}}
\newcommand{\brmC}{\mbox{\bf{C}}}
\newcommand{\brmW}{\mbox{\bf W}}
\newcommand{\bU}{\mbox{\boldmath{$U$}}}

\newcommand{\bh}{\mbox{\boldmath{$h$}}}
\newcommand{\bb}{\mbox{\boldmath{$b$}}}
\newcommand{\bs}{\mbox{\boldmath{$s$}}}

\newcommand{\bm}{\mbox{\boldmath{$m$}}}
\newcommand{\by}{\mbox{\boldmath{$y$}}}

\begin{document}
\baselineskip 0mm

\vspace*{1cm}
\begin{center}{\LARGE \bf 
A statistical-mechanical approach to CDMA multiuser detection:
propagating beliefs in a densely connected graph 
} 
\end{center}
\begin{center}
{\large Yoshiyuki Kabashima\\
Department of Computational Intelligence \& Systems Science \\
Tokyo Institute of Technology, Yokohama 2268502, Japan. \\
}
\end{center}


\begin{abstract}
The task of CDMA multiuser detection is to simultaneously estimate 
binary symbols of $K$ synchronous users from the received $N$ base-band 
CDMA signals. Mathematically, this can be formulated as an inference 
problem on a complete bipartite graph. 
In the research on graphically represented statistical models, 
it is known that the belief propagation (BP) can 
exactly perform the inference in a polynomial time scale 
of the system size when the graph is free from cycles 
in spite that the necessary computation for general graphs
exponentially explodes in the worst case \cite{Pearl}. 
In addition, recent several researches revealed that 
BP can also serve as an excellent approximation algorithm 
even if the graph has cycles as far as they are relatively 
long \cite{MacKay,Kaba,Yedidia}. However, as there exit many short cycles 
in a complete bipartite graph, one might suspect that the 
BP would not provide a good performance when employed 
for the multiuser detection. 

The purpose of this paper is to make an objection to such suspicion. 
More specifically, we will show that appropriate 
employment of the central limit theorem and the law 
of large numbers to BP, which is
one of the standard techniques in statistical mechanics, 
makes it possible to 
develop a novel multiuser detection algorithm 
the convergence property of which is considerably 
better than that of the conventional multistage 
detection \cite{multistage} without increasing the computational 
cost significantly. Furthermore, we will also provide 
a scheme to analyse the dynamics of 
the proposed algorithm, which can be naturally linked to the 
equilibrium analysis recently presented by Tanaka in \cite{tanaka}. 
\end{abstract}

\section{Multiuser detection}
We will focus on a CDMA system using binary shift keying 
(BPSK) symbols and $K$ random binary spreading codes of 
the spreading factor $N$ with unit energy over 
an additive white Gaussian noise (AWGN) channel. 
For simplicity, we assume the power is completely controlled to 
unit energy; 
but the extension to the case of distributed powers is straightforward. 
Under these assumptions, a received base-band CDMA signal 
is expressed as
\begin{eqnarray}
y_\mu=\frac{1}{\sqrt{N}}\sum_{k=1}^Ks_{\mu k} b_k + \sigma_0 n_\mu,  
\label{eq:base_band}
\end{eqnarray}
where $\mu \in \{1,2,\ldots,N\}$ and $k \in \{
1,2,\ldots,K\}$ are indices for samples and users, respectively. 
$s_{\mu k} \in \{-1,1\}$ is the spreading code with unit energy independently 
generated from the identical unbiased distribution 
$P(s_{\mu k}=+1)=P(s_{\mu k}=-1)=1/2$ and $b_k$ is 
the bit signal of user $k$. $n_\mu$ is a Gaussian white noise 
sample with zero mean and unit variance and $\sigma_0$ 
is the standard deviation of AWGN. 
Using these normalisations, the signal to noise ratio is defined 
as $SNR=\beta/(2 \sigma_0^2)$ where $\beta = K/N$. 
In the following, we assume a situation where both of $N$ and $K$ are large
keeping $\beta$ finite. 

The goal of multiuser detection is to simultaneously infer the bit 
signals $b_1, b_2, \ldots,b_K$ after receiving the base-band 
signals $y_1,y_2,\ldots,y_N$. 
The Bayesian approach offers a useful framework for such purposes. 
Assuming that the bit signals are independently generated 
from the unbiased distribution, the posterior distribution 
given the base-band signals is provided as
\begin{eqnarray}
P(\bb|\by)=\frac{\prod_{\mu=1}^N P(y_\mu|\bb)}
{\sum_{\bb} \prod_{\mu=1}^N P(y_\mu|\bb)}, 
\label{eq:posterior}
\end{eqnarray} 
where
\begin{eqnarray}
P(y_\mu|\bb) = \frac{1}{\sqrt{2 \pi \sigma_0^2}}
\exp \left [ -\frac{1}{2 \sigma_0^2} 
\left (y_\mu - \Delta_\mu \right )^2 \right ], \label{eq:conditional}
\end{eqnarray}
and $\Delta_\mu \equiv \frac{1}{\sqrt{N}} \sum_{k=1}^K s_{\mu k}b_k$. 
Following the Bayesian framework, one can systematically 
derive the optimal inference strategy from the 
posterior distribution (\ref{eq:posterior}) for various cost functions. 
For instance, it can be shown that the bit error 
rate (BER), which is the cost function that we will 
focus on in this paper, is minimised 
by the maximiser of the posterior marginal (MPM) estimator
\begin{eqnarray}
\hat{b}_k=\mathop{\rm argmax}_{b_k \in \{+1,-1\}}
\sum_{b_{l \ne k}} P(\bb|\by). 
\label{eq:MAP}
\end{eqnarray}

\section{Graphical expression and belief propagation}
Unfortunately, the necessary cost for exactly computing the MPM estimator 
explodes exponentially with respect to the number of users $K$ in the 
current system, which implies that one has to resort to an approximation 
in practice. The belief propagation (BP), or 
the sum-product algorithm, is known as 
one of the most promising approaches to such tasks 
although its performance for densely connected systems, 
including complete bipartite graphs, has not been 
sufficiently examined yet \cite{Opper_Saad}. 
We here investigate the efficacy of BP in densely connected systems
employing it to the present CDMA multiuser detection problem. 

\begin{figure}[t]
\setlength{\unitlength}{1mm}
\epsfxsize = 70mm
\begin{center}
\epsfbox{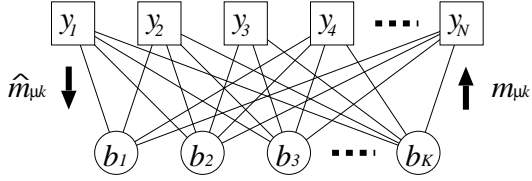}
\end{center}
\vspace*{-.5cm}
\caption{Graphical expression of the CDMA multiuser detection problem. 
Each edge corresponds to a component of spreading codes $s_{\mu k}$. }
\label{fig:bipartite}
\end{figure}

In order to introduce this algorithm to the current system, 
let us denote the base-band and bit signals by two kinds of nodes 
and connect them with an edge when they are related. 
Since the conditional probability of $y_\mu$ (\ref{eq:conditional}) 
depends on all of $b_1,b_2,\ldots,b_k$,
this implies that the posterior distribution (\ref{eq:posterior}) can be
expressed as a complete bipartite graph as shown in Figure 
\ref{fig:bipartite}. 

Then, BP can be defined as an algorithm 
passing messages between the two kinds of nodes through edges as 
\begin{eqnarray}
P^{t+1}\left (y_{\mu}| b_k, \{y_{\nu \ne \mu }\} \right ) &\propto&
\hat{\alpha}_{\mu k}^{t+1} \sum_{b_{l \ne k}} P(y_\mu|\bb)\prod_{l \ne k }
P^t(b_l|\{y_{\nu \ne \mu}\}), \label{eq:horizontal} \\
P^t\left (b_k|\{y_{\nu \ne \mu }\} \right ) &= &\alpha_{\mu k}^t
\prod_{\nu \ne \mu} P^t\left (y_\nu| b_k, \{y_{\sigma \ne \nu }\} \right ),
\label{eq:vertical}
\end{eqnarray}
where $t=1,2, \ldots$ is an index for counting the number of updates, 
$\hat{\alpha}_{\mu k }^t$ and $\alpha_{\mu k}^t$ are
constants for normalisation constraints $\sum_{b_k=\pm 1}
P^t\left (y_{\mu}| b_k, \{y_{\nu \ne \mu }\} \right ) =1$ 
and $\sum_{b_k=\pm 1} P^t\left (b_k|\{y_{\nu \ne \mu }\} \right )=1$, 
respectively. 
The 
marginalised posterior at $t$th update is evaluated
from $P^t\left (y_{\mu}| b_k, \{y_{\nu \ne \mu }\} \right )$ as
$P^t(b_k|\by)=\alpha_k \prod_{\mu=1}^N 
P^t\left (y_{\mu}| b_k, \{y_{\nu \ne \mu }\} \right )$, 
where $\alpha_k$ is a normalisation constant. 

As $b_k$ is a binary variable, one can parameterise the above functions as 
$P^t\left (y_{\mu}| b_k, \{y_{\nu \ne \mu }\} \right )
\propto (1+\hat{m}^t_{\mu k}b_k  )/2$, 
$P^t\left (b_k|\{y_{\nu \ne \mu }\} \right )=
 (1+m^t_{\mu k} b_k  )/2$ 
and $P^t(b_k|\by)=(1+m_k^t b_k  )/2$
without loss of generality, which simplifies the expressions
(\ref{eq:horizontal}) and (\ref{eq:vertical}) as
\begin{eqnarray}
\hat{m}_{\mu k}^{t+1}&=&\frac{\sum_{\bb} 
b_k P(y_\mu|\bb)
\prod_{l \ne k} \left (
\frac{1+m_{\mu l}^t b_l}{2} \right )}
{\sum_{\bb} 
P(y_\mu|\bb) \prod_{l \ne k} \left (
\frac{1+m_{\mu l}^t b_l}{2} \right )}, \label{eq:hat} \\
m_{\mu k}^t&=&\tanh\left (\sum_{\nu \ne \mu} 
\tanh^{-1} \hat{m}_{\nu k}^t \right ). 
\label{eq:nohat} 
\end{eqnarray}
Employing these variables, the approximated posterior average 
of $b_k$ at $t$th update 
can be computed as $m_{k}^t = \tanh\left (\sum_{\mu=1}^N 
\tanh^{-1} \hat{m}_{\mu k}^t \right )$. 

\section{Propagating beliefs in a large complete bipartite graph} 
Reflecting the fact that each base-band signal $y_\mu$ is connected 
with every bit signal $b_k$, evaluating eq. (\ref{eq:hat}) 
brings about a computational explosion when $K$ is large, which implies 
exactly performing BP becomes hopeless in the current system. 
However, appropriately 
employing the central limit theorem 
and the law of large numbers, which 
is a standard procedure in 
statistical-mechanical analysis 
\cite{Opper_Winther,Kaba_Saad}, makes it possible to 
approximately carry out the belief updates (\ref{eq:hat}) and 
(\ref{eq:nohat}) in a practical time scale. 

Since $s_{\mu k}b_k/\sqrt{N}$ is small for large $N$, we expand 
the conditional probability as
\begin{eqnarray}
P(y_\mu|\bb) &\simeq& \frac{1}{\sqrt{2\pi \sigma_0^2}}
\exp\left [ -\frac{\left (y_\mu -\Delta_{\mu k} \right)^2
}{2 \sigma_0^2}+\frac{s_{\mu k} (y_\mu-\Delta_{\mu k})}{\sqrt{N} \sigma_0^2}
b_k \right ] 
\cr 
&\simeq&\frac{1}{\sqrt{2\pi \sigma_0^2}}
\exp\left [ -\frac{\left (y_\mu -\Delta_{\mu k} \right)^2}{2 \sigma_0^2 }
\right]\left (1+ \frac{s_{\mu k} 
(y_\mu-\Delta_{\mu k})}{\sqrt{N} \sigma_0^2} b_k \right ), 
\label{expansion}
\end{eqnarray}
where $\Delta_{\mu k} \equiv \sum_{l \ne k} s_{\mu l} b_l/\sqrt{N}$ 
in eq. (\ref{eq:hat}). As the spreading codes are generated independently, 
$s_{\mu l}$ and $b_l$ would be uncorrelated when $b_l$ is generated 
from $P^t(b_l|\{y_{\nu \ne \mu}\})=(1+m_{\mu l}^tb_l)/2$.
This, in conjunction with the central limit theorem, implies 
that $\Delta_{\mu k} \equiv \sum_{l \ne k} s_{\mu l} b_l/\sqrt{N}$ obeys
a normal distribution ${\cal N}\left (\left \langle 
\Delta_{\mu k}^t \right \rangle_{\mu},\beta (1-Q_{\mu k}^t ) \right )$, 
where $\left \langle 
\Delta_{\mu k}^t \right \rangle_{\mu} \equiv 
\sum_{l \ne k }s_{\mu l} m_{\mu l}^t /\sqrt{N} $ and 
$Q^t_{\mu k}\equiv (1/K) \sum_{l \ne k} (m_{\mu l}^t)^2$. 
Furthermore, due to the law of large numbers, 
$Q_{\mu k}^t$ is highly likely to be well approximated by 
$Q^t \equiv (1/K) \sum_{k=1}^K (m_l^t)^2$. 
Substituting these, one can evaluate eq. (\ref{eq:hat}) as
\begin{eqnarray}
\hat{m}_{\mu k}^{t+1}
=
A^t\left (\frac{y_\mu \bs_\mu}{\sqrt{N}} -\beta \left (\brmP_\mu -
\frac{\brmI}{K} \right ) \bm_\mu^t \right )_k, 
\label{eq:simplified_1}
\end{eqnarray}
where $\bs_\mu \equiv (s_{\mu k})$, $\bm_{\mu}^t \equiv (m_{\mu k}^t)$ 
and $A^t\equiv 
\left (\sigma_0^2+\beta(1-Q^t) \right )^{-1}$. 
Here, we also introduced the projection and the identity matrices 
$\brmP_\mu \equiv (1/K) (s_{\mu k} s_{\mu l})$ and $\brmI 
\equiv (\delta_{kl})$, respectively. 
$(\cdots)_k$ denotes $k$th component of the vector $\cdots$. 
Eq. (\ref{eq:simplified_1}) can be evaluated by $O(K)$ computations 
per pair $(\mu k)$, which implies that $O(NK^2)$ 
computations are totally required per update. 

The computational cost can be further reduced to $O(K^2)$ 
when $N$ is large 
employing eq. (\ref{eq:nohat}). As $\hat{m}_{\mu k}^t$ typically 
scales as $O(N^{-1/2})$, eq. (\ref{eq:nohat}) can be 
expanded as
$m_{\mu k}^t \simeq m_k^t - (\partial m_k/\partial \hat{m}_{\mu k}^{t}) 
\hat{m}_{\mu k}^t =m_k^t - (1-\left (m_k^t)^2 \right ) \hat{m}_{\mu k}^t$. 
Plugging this into eq. (\ref{eq:simplified_1}) provides 
a recursive equation with respect to 
$\hat{\bm}_\mu^t \equiv (\hat{m}_{\mu k}^t)$ as
\begin{eqnarray}
\hat{\bm}_\mu^{t+1} = A^t \frac{y_\mu \bs_\mu}{\sqrt{N}} -
\beta A^t \left (\brmP_\mu -
\frac{\brmI}{K} \right ) \bm^t + \beta A^t \brmP_\mu \brmC^t \hat{\bm}_\mu^t, 
\label{eq:recursive}
\end{eqnarray}
where $\brmC^t \equiv ((1-(m_k^t)^2) \delta_{kl})$. 
Employing useful relations $\brmP_\mu \brmC^t \bs_\mu=
(1-Q^t) \bs_\mu$ 
and $\brmP_\mu \brmC^t \brmP_\mu =(1-Q^t) \brmP_\mu$ and 
omitting negligible terms, the solution of 
eq. (\ref{eq:recursive}) can be expressed as
\begin{eqnarray}
\hat{\bm}_\mu^{t+1}=R^t \frac{y_\mu \bs_\mu}{\sqrt{N}} -
\bU^t_\mu+\frac{1}{K}\beta A^t \bm^t , 
\end{eqnarray}
where $R^t$ and $\bU^t$ are obtained from recursive equations
\begin{eqnarray}
R^{t}&=&A^t + A^t\beta(1-Q^t) R^{t-1}, \label{eq:Rt_update} \\
\bU^{t}_\mu &=& A^t \beta \brmP_\mu \bm^t +A^t \beta (1-Q^t) \bU^{t-1}_\mu. 
\label{eq:Ut_update}
\end{eqnarray}
Since $\hat{m}_{\mu k}$ typically scales as $O(N^{-1/2})$, 
the posterior average can be expressed as 
$m_{k}^t = \tanh\left (\sum_{\mu=1}^N \tanh^{-1} \hat{m}_{\mu k}^t \right )
\simeq \tanh\left (\sum_{\mu=1}^N \hat{m}_{\mu k}^t \right )$. 
This implies that the belief updates (\ref{eq:horizontal})
and (\ref{eq:vertical}) are finally summarised into 
\begin{eqnarray}
\bh^{t+1}&=&R^t \bh^0 -\bU^t + A^t \bm^t, \label{eq:horizontal2} \\
\bU^t&=&A^t \beta \brmW \bm^t+A^t \beta(1-Q^t) \bU^{t-1}, \label{eq:Uupdate}
\end{eqnarray}
and eq. (\ref{eq:Rt_update}), where $m_k^t =\tanh(h_k^t)$, 
$\bh^0 \equiv (h_k^0)\equiv(\sum_{\mu=1}^N y_\mu s_{\mu k}/\sqrt{N})$, 
$\bh^t\equiv (h_k^t)$ 
and $\brmW \equiv (W_{kl})\equiv\left (\sum_{\mu=1}^N s_{\mu k}s_{\mu l}/N 
\right )$. From the posterior average $m_k^t$, 
the MPM estimator at $t$th update 
is evaluated as $\hat{b}_k^t=\sign(m_k^t)$ where $\sign(x)\equiv 
\lim_{\epsilon \to +0} x/|x+\epsilon|$. 

Two points are worthy of noticing. Firstly, the most time-consuming 
operation in eqs. (\ref{eq:Rt_update}), (\ref{eq:horizontal2})
and (\ref{eq:Uupdate}) is $\brmW \bm^t$, which totally requires
$O(K^2)$ computations. This implies that the computational cost 
for performing the current scheme is similar to that of the 
conventional multistage detection \cite{multistage}
\begin{eqnarray}
\hat{b}_{k}^{t+1} = \sign \left (h_k^0-\sum_{ l \ne k } W_{kl} \hat{b}_l^t
\right ). 
\label{eq:multistage}
\end{eqnarray}
Secondly, as the fixed point condition, coupled nonlinear equations 
\begin{eqnarray}
m_k=\tanh \left [\sigma_0^{-2} \left (h_k^0 -
\sum_{l \ne k} W_{kl} m_l \right )-
\frac{\beta (1-Q) m_k}{\sigma_0^2 (\sigma_0^2+\beta
(1-Q))} \right ], 
\label{eq:TAP}
\end{eqnarray}
are obtained from our update scheme, where $Q=(1/K) \sum_{k=1}^k m_l^2$. 
This is identical to the Thouless-Anderson-Palmer (TAP) 
equation for the current system known in statistical 
mechanics \cite{Opper}\footnote{
In statistical physics, {\em pattern ratio }
$\tilde{\alpha}=N/K=\beta^{-1}$ and {\em inverse temperature} 
$\tilde{\beta}=\beta \sigma_0^{-2}$ are usually employed for 
characterising a system instead of $\beta$ and $\sigma_0^2$. }.
However, it should be emphasised here that, the naive iteration of 
eq. (\ref{eq:TAP}) does not serve as a useful detection 
algorithm as finding the fixed point by it
from a reasonable initial state is difficult. 
This will be illustrated by numerical experiments in the final section. 

\section{Density evolution and equilibrium analysis}
The density evolution is a framework to analyse the dynamical 
property of BP pursuing a macroscopic distribution 
of messages \cite{Richardson,TanakaISIT2002}. 
In the current system, this analysis is considerably simplified 
as the aligned field $b_k h_k^t$ is likely to obey a normal 
distribution as a result of the central limit theorem. 

Let us assume that $b_k h_k^t=b_k \sum_{\mu=1}^N\hat{m}_{\mu k}^t$ 
is independently sampled from a normal distribution the 
average and variance of which 
are $E^t$ and variance $F^t$, respectively. This implies that 
the overlap $M^t\equiv \sum_{k=1}^K b_k m_k^t/K$ and $Q^t$ are evaluated as
\begin{eqnarray}
M^t=\int Dz \tanh (\sqrt{F^t}z +E^t), \quad 
Q^t=\int Dz \tanh^2(\sqrt{F^t}z +E^t),
\label{eq:EF_QM}
\end{eqnarray}
where $Dz\equiv dz \exp[-z^2/2]/\sqrt{2 \pi}$. 
Since the MPM estimator is given as $\hat{b}_k^t=\sign(h^t_k)$, 
BER is provided as $P_b^t=(1/K) \sum_{k=1}^K {(1-\sign(b_k h_k^t))/2}=
\int_{-\infty}^{-E^t/\sqrt{F^t}} Dz$. 

On the other hand, as far as $b_k h_k^t$ is independently sampled, 
$b_k \hat{m}_{\mu k}^{t+1}$ evaluated from eq. (\ref{eq:simplified_1})
is uncorrelated for a given $k$ since spreading codes $\bs_\mu$ 
are almost orthogonal with each other when $N$ is large
as $\bs_\mu \cdot \bs_\nu /N \simeq O(N^{-1/2})$ holds for 
$\mu \ne \nu$. 
This implies that the central limit theorem holds for $b_k h_k^{t+1}$, 
which, in conjunction with the statistical 
uniformness with respect to indices $\mu$ and $k$, 
provides the average and the variance at $t+1$st update as 
$E^{t+1}=(1/K)\sum_{k=1}^K \sum_{\mu=1}^N 
b_k \hat{m}_{\mu k}^{t+1} =(1/K)\sum_{\mu=1}^N \bb\cdot \hat{\bm}_\mu^t$
and 
$F^{t+1}=(1/K)\sum_{k=1}^K \sum_{\mu=1}^N \left (b_k \hat{m}_{\mu k}^t 
\right )^2 =(1/K)\sum_{\mu=1}^N \hat{\bm}_\mu^t \cdot \hat{\bm}_\mu^t$, 
respectively. Evaluating these employing eqs. (\ref{eq:base_band})
and (\ref{eq:simplified_1}), $E^{t+1}$ and 
$F^{t+1}$ are obtained as
\begin{eqnarray}
E^{t+1}=\frac{1}{\sigma_0^2+\beta(1-Q^t)}, \quad
F^{t+1}=\frac{\beta (1-2M^t+Q^t)+\sigma_0^2}{
\left [\sigma_0^2+\beta(1-Q^t) \right ]^2},
\label{eq:MQ_EF}
\end{eqnarray}
where we assumed that $(1/K) \sum_{k=1}^K \bb \cdot \bm_\mu^t
\simeq M^t$ holds as a result from the law of large numbers. 
Eqs. (\ref{eq:EF_QM}) and (\ref{eq:MQ_EF}) express the density 
evolution with respect to the current algorithm. 

It should be noticed that the obtained expression of the density evolution 
directly links the proposed algorithm to the equilibrium analysis 
presented in \cite{tanaka} since eqs. (\ref{eq:EF_QM}) and 
(\ref{eq:MQ_EF}) can be regarded as the naive iteration dynamics 
of the saddle point equations provided by the replica method \cite{replica}. 
This implies that our algorithm can practically 
calculate the MPM estimator (\ref{eq:MAP}) in $O(K^2)$ computations
obtaining the fixed point solution when $K$ is large 
since the replica analysis is likely to evaluate the exact 
performance for $K \to \infty$, which, however, 
has not been rigorously proved yet.

\section{Method comparison and discussion}
\begin{figure}[t]
\setlength{\unitlength}{1mm}
\epsfxsize = 100mm
\begin{center}
\epsfbox{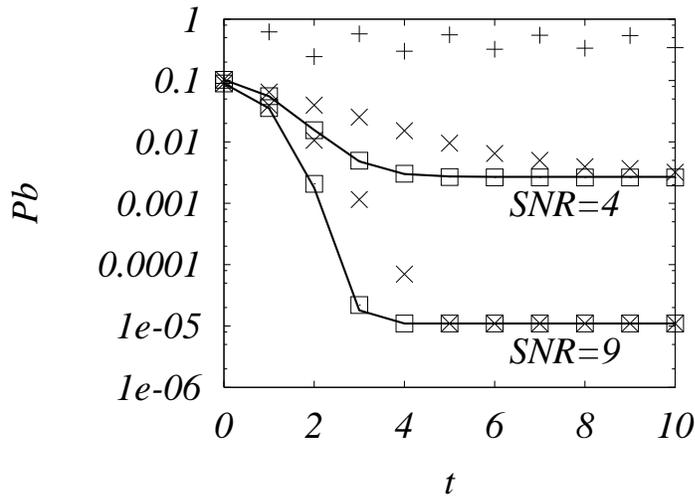}
\end{center}
\vspace*{-.5cm}
\caption{
Time evolution of BER for the proposed algorithm (PA:$\Box$), 
the conventional multistage detection (MSD:$\times$), 
the naive iteration of the TAP equation (TAP:$+$)
and the density evolution (DE: lines)
in the case of $N=2000$, $\beta=0.5$ and $SNR=4,9$
(data of TAP are shown only for $SNR=9$). 
Each marker represents the averaged BER at $t$th update 
evaluated from $10000$ experiments. 
PA exhibits the fastest convergence and excellent consistency with 
DE.}
\label{fig:evolution}
\end{figure}

In order to validate the obtained results, we performed 
numerical experiments in a system $N=2000$ and $\beta=0.5$. 
Figure \ref{fig:evolution} shows time evolution of BER obtained 
from $10000$ experiments for the proposed algorithm 
(eqs. (\ref{eq:Rt_update}), (\ref{eq:horizontal2}) 
and (\ref{eq:Uupdate}): PA), 
the conventional multistage detection 
(eq. (\ref{eq:multistage}): MSD), 
the iteration of the TAP equation 
(eq. (\ref{eq:TAP}): TAP)
and the density evolution (eqs. (\ref{eq:EF_QM}) and (\ref{eq:MQ_EF}): DE).

Firstly, it is clear that PA converges to the fixed point considerably 
faster than MSD, which is a highly preferred property in practical use. 
Secondly, PA and DE exhibit excellent 
consistency as we speculated in the previous section, which implies 
that employment of the central limit theorem and the law of 
large numbers for deriving eqs. (\ref{eq:EF_QM}) and 
(\ref{eq:MQ_EF}) is fully validated. 
Finally, TAP does not serve as a useful detection algorithm. 
This is because the iteration of eq. (\ref{eq:TAP}) does not 
correctly approximate BP and, therefore, 
can not sufficiently cancel self-reactions from the past states 
in the transient dynamics although it does provide the correct 
fixed point condition in the stationary state. 

In summary, we have developed a novel algorithm for the 
CDMA multiuser detection from the belief propagation 
appropriately employing the central limit theorem
and the law of large numbers. 
The new algorithm exhibits considerably 
faster convergence than the conventional multistage detection 
without increasing the computational cost significantly 
and is likely to practically provide the optimal 
MPM estimator when the spreading factor $N$ is large. 
We have also clarified the relation between the obtained algorithm
and the existing equilibrium analysis presented in \cite{tanaka}
employing the density evolution scheme. 

We have here assumed randomly generated spreading codes 
for simplicity, which might not be suitable for practical use. 
Extension of the current scheme to other methods of code generation 
is under way. 

\section*{Acknowledgements}
The author thanks Toshiyuki Tanaka for his useful discussion. 
Support from Grant-in-Aid, MEXT, Japan, Nos. 13680400, 13780208 and 14084206
is acknowledged.

\end{document}